\documentclass[twocolumn,preprintnumbers,amsmath,amssymb,aps,superscriptaddress]{
revtex4}
\usepackage{amsfonts}
\usepackage{amsmath}
\usepackage{amssymb}
\usepackage{graphicx,color}
\usepackage{rotating}
\usepackage{dcolumn}
\usepackage{bm,ulem}
\usepackage[colorlinks=true, pdfstartview=FitV, linkcolor=blue, 
citecolor=blue,urlcolor=navyblue]{hyperref}

\begin{document}

\title{Consistent relativistic mean-field models: critical parameter values} 

\author{M. Dutra$^*$}
\affiliation{Departamento de Ci\^encias da Natureza - IHS, Universidade Federal
Fluminense, 28895-532 Rio das Ostras, RJ, Brazil}

\author{O. Louren\c{c}o \footnote{Present address: Departamento de F\'isica, Instituto 
Tecnol\'ogico de Aeron\'autica, CTA, S\~ao Jos\'e dos Campos, 12228-900. SP,
Brazil}}
\affiliation{Universidade Federal do Rio de Janeiro, 27930-560, Maca\'e, RJ, Brazil}

\author{D. P. Menezes} 
\affiliation{Depto de F\'{\i}sica - CFM - Universidade Federal de Santa Catarina, Florian\'opolis 
- SC - CP. 476 - CEP 88.040 - 900 - Brazil}

\date{\today}

\begin{abstract}
We revisit the study published in \cite{prc-critical}, related to the behavior of $34$ relativistic 
mean-field models, previously selected because they satisfy bulk nuclear matter properties, here 
used to compute the critical parameters of the symmetric nuclear matter. We evaluate their critical 
temperature, pressure, and density and compare with some values encountered in the literature. We 
also show that these parameters are correlated with the incompressibility calculated at the zero 
temperature regime.
\end{abstract}

\maketitle

\section{Introduction}

Nuclear matter is an idealized medium, and all its properties, are derived from experiments 
indirectly in a model-dependent way. However, the understanding of its properties is of fundamental 
importance as a guide towards more specific subjects, such as nuclear and hadron spectroscopy, 
heavy-ion collisions, nuclear multifragmentation, caloric curves, and others. It is well known that 
theoretical hadronic models predict phase transitions at moderate temperatures. All these models 
share the prediction that a liquid-gas phase transition occurs for symmetric and asymmetric nuclear 
matter at finite temperature and density. Qualitatively, the isotherms of these hadronic mean-field 
models typically show a van der Waals-like behavior, where liquid and gaseous phases can coexist 
\cite{vdw5}. 

It is important to note that the critical temperature is defined in such a way that it always takes 
place in symmetric matter. In Ref.\cite{avancini2006} the authors have shown that the instability 
region decreases with the increase of the temperature up to a certain value, which is related to a 
critical pressure and critical density. The values of these critical parameters are model dependent 
and there are many nonrelativistic and relativistic models in the  literature, which can be used to 
calculate them. In this work we use the Relativistic Mean-Field (RMF) Approximation.

\section{Choice of models}

Our study is based on the study presented in Ref.~\cite{rmf} where $263$ RMF models were analyzed. 
These parametrizations had their volumetric and thermodynamical quantities compared with theoretical 
and experimental data available in the literature. These data were divided into three groups: 
symmetric nuclear matter (SNM), pure neutron matter (PNM) and a third group named MIX (the mixture 
of PNM + SNM). This last one encompasses the symmetry energy and its slope at the saturation density 
as well as reduction of the symmetry energy at half of the saturation density. In Table~\ref{set2a} 
we present a summary of these constraints. For more details see Ref.~\cite{rmf}. 

The analysis has shown that only 35 parametrizations were approved. They are named consistent 
relativistic mean field (CRMF) parametrizations. We consider in the present study only $34$ of 
them because the point-coupling parametrization does not generate a mass-radius curve, according to 
Ref.~\cite{stars}, so, it was excluded. The remaining of them are part of two groups out of the 
seven presented in Ref.~\cite{rmf}. We have shown them in Subsection ~\ref{seca} and \ref{secb} of 
that reference".

\subsection{Nonlinear RMF models\label{seca}}

The group of the nonlinear RMF parametrizations with $\sigma$ and $\omega$ terms 
and cross terms involving these fields encompasses thirty parametrizations. The 
Lagrangian density that describes this model is:

\begin{eqnarray}
\mathcal{L}_{\rm NL} &=& \overline{\psi}(i\gamma^\mu\partial_\mu - M)\psi 
+ g_\sigma\sigma\overline{\psi}\psi - g_\omega\overline{\psi}\gamma^\mu\omega_\mu\psi 
\nonumber \\
&-& \frac{g_\rho}{2}\overline{\psi}\gamma^\mu\vec{\rho}_\mu\vec{\tau}\psi
+ \frac{1}{2}(\partial^\mu \sigma \partial_\mu \sigma 
- m^2_\sigma\sigma^2) - \frac{A}{3}\sigma^3 
\nonumber\\
&-&  \frac{B}{4}\sigma^4 -\frac{1}{4}F^{\mu\nu}F_{\mu\nu} 
+ \frac{1}{2}m^2_\omega\omega_\mu\omega^\mu + 
\frac{C}{4}(g_\omega^2\omega_\mu\omega^\mu)^2 
\nonumber \\
&-& \frac{1}{4}\vec{B}^{\mu\nu}\vec{B}_{\mu\nu} + 
\frac{1}{2}m^2_\rho\vec{\rho}_\mu\vec{\rho}^\mu
+ \frac{1}{2}{\alpha_3'}g_\omega^2
g_\rho^2\omega_\mu\omega^\mu\vec{\rho}_\mu\vec{\rho}^\mu
\nonumber\\
&+& g_\sigma g_\omega^2\sigma\omega_\mu\omega^\mu
\left(\alpha_1+\frac{1}{2}{\alpha_1'}g_\sigma\sigma\right)
\nonumber\\
&+& g_\sigma g_\rho^2\sigma\vec{\rho}_\mu\vec{\rho}^\mu
\left(\alpha_2+\frac{1}{2}{\alpha_2'}g_\sigma\sigma\right),
\label{lomegarho}
\end{eqnarray}
with $F_{\mu\nu}=\partial_\nu\omega_\mu-\partial_\mu\omega_\nu$
and $\vec{B}_{\mu\nu}=\partial_\nu\vec{\rho}_\mu-\partial_\mu\vec{\rho}_\nu$. The nucleon 
mass is $M$ and the meson masses are $m_\sigma$,$m_\omega$, and $m_\rho$.

\begin{table}[htb!]
\begin{ruledtabular}
\caption{Set of updated constraints (SET2a) used in Ref.~\cite{rmf}. For more 
details concerning each constraint see the reference.}
\centering
\begin{tabular}{lccc}
Constraint & Quantity      & Density Region       & Range of constraint \\ 
\hline
SM1    & $K_0$     & at $\rho_0$  & 190 $-$ 270 MeV \\
SM3a   & $P(\rho)$ & $2<\frac{\rho}{\rho_0}<5$ & Band Region \\
SM4    & $P(\rho)$ & $1.2<\frac{\rho}{\rho_0}<2.2$    & Band Region\\
PNM1   & $\mathcal{E}_{\mbox{\tiny PNM}}/\rho$ & $0.017<\frac{\rho}{\rho_{\rm
o}}<0.108$   & Band Region \\
MIX1a  & $J$       & at $\rho_0$  & 25 $-$ 35 MeV \\
MIX2a  & $L_0$     & at $\rho_0$  & 25 $-$ 115 MeV \\
MIX4   & $\frac{\mathcal{S}(\rho_0/2)}{J}$  & at $\rho_0$ and $\rho_0/2$ & 0.57 $-$ 0.86\\
\end{tabular}
\label{set2a}
\end{ruledtabular}
\end{table}

We can derive from Eq.(\ref{lomegarho}) the equation of state for 
symmetric nuclear matter ($\gamma=4$). The pressure is given by
\begin{align}
P_{\rm NL} &= - \frac{1}{2}m^2_\sigma\sigma^2 - \frac{A}{3}\sigma^3 -
\frac{B}{4}\sigma^4 + \frac{1}{2}m^2_\omega\omega_0^2 
+ \frac{C}{4}(g_\omega^2\omega_0^2)^2
\nonumber\\
+& g_\sigma g_\omega^2\sigma\omega_0^2
\left(\alpha_1+\frac{1}{2}{\alpha_1'}g_\sigma\sigma\right) 
\nonumber\\
+& \dfrac{\gamma}{6\pi^2}\int_0^{\infty}\dfrac{dk\,k^4}{(k^2 + 
{M^*}^2)^{1/2}}\left[n(k,T,\mu^*)+\bar{n}(k,T,\mu^*)\right],\nonumber\\
\label{pnl}
\end{align}
where
\begin{align}
n(k,T,\mu^*) &= \frac{1}{e^{(E^*-\mu^*)/T}+1},\quad\mbox{and}\nonumber\\
\bar{n}(k,T,\mu^*) &= \frac{1}{e^{(E^*+\mu^*)/T}+1}
\label{fermi-dirac}
\end{align}
are the Fermi-Dirac distributions for particles and antiparticles, respectively. The 
effective energy, nucleon mass, and chemical potential are $E^*=(k^2+{M^*}^2)^{1/2}$, 
$M^*=M-g_\sigma\sigma$, and $\mu^*=\mu-g_\omega\omega_0$, respectively. Furthermore, the 
(classical) mean-field values of $\sigma$ and $\omega_0$ are found by solving the 
following system of equations,
\begin{eqnarray}
m^2_\sigma\sigma &=& g_\sigma\rho_s - A\sigma^2 - B\sigma^3 
+g_\sigma g_\omega^2\omega_0^2(\alpha_1+{\alpha_1'}g_\sigma\sigma)
\\
m_\omega^2\omega_0 &=& g_\omega\rho - Cg_\omega(g_\omega \omega_0)^3 
- g_\sigma g_\omega^2\sigma\omega_0(2\alpha_1+{\alpha_1'}g_\sigma\sigma),
\nonumber\\
\end{eqnarray}
with
\begin{align}
\rho&=\dfrac{\gamma}{2\pi^2}\int_0^{\infty}dk\,k^2
\left[n(k,T,\mu^*)-\bar{n}(k,T,\mu^*)\right],
\label{rho}
\\
\rho_s &= \dfrac{\gamma}{2\pi^2}\int_0^{\infty}\frac{dk\,M^*k^2}{(k^2+{M^*}^2)^{1/2}}
\left[n(k,T,\mu^*)+\bar{n}(k,T,\mu^*)\right].
\label{rhos}
\end{align}

\subsection{Density-dependent models\label{secb}}

The four remaining parametrizations belong to the density-dependent group. Two of them include 
the $\delta$ meson. Their Lagrangian density is given by
\begin{eqnarray}
\mathcal{L}_{\rm DD} &=& \overline{\psi}(i\gamma^\mu\partial_\mu - M)\psi 
+ \Gamma_\sigma(\rho)\sigma\overline{\psi}\psi 
- \Gamma_\omega(\rho)\overline{\psi}\gamma^\mu\omega_\mu\psi 
\nonumber\\
&-&\frac{\Gamma_\rho(\rho)}{2}\overline{\psi}\gamma^\mu\vec{\rho}_\mu\vec{\tau}
\psi + \Gamma_\delta(\rho)\overline{\psi}\vec{\delta}\vec{\tau}\psi 
- \frac{1}{4}F^{\mu\nu}F_{\mu\nu}
\nonumber \\
&+& \frac{1}{2}(\partial^\mu \sigma \partial_\mu \sigma - m^2_\sigma\sigma^2)
 + \frac{1}{2}m^2_\omega\omega_\mu\omega^\mu 
-\frac{1}{4}\vec{B}^{\mu\nu}\vec{B}_{\mu\nu}
\nonumber \\
&+& \frac{1}{2}m^2_\rho\vec{\rho}_\mu\vec{\rho}^\mu + 
\frac{1}{2}(\partial^\mu\vec{\delta}\partial_\mu\vec{\delta} 
- m^2_\delta\vec{\delta}^2),
\label{dldd}
\end{eqnarray}
where
\begin{eqnarray}
\Gamma_i(\rho) &=& \Gamma_i(\rho_0)f_i(x);\quad
f_i(x) = a_i\frac{1+b_i(x+d_i)^2}{1+c_i(x+d_i)^2},
\label{gamadefault}
\end{eqnarray}
for $i=\sigma,\omega$, and $x=\rho/\rho_0$. 

The expression for the pressure for these models can be obtained from
Eq.~(\ref{dldd}) and reads:
 
\begin{align}
P_{\rm DD} &= \rho\Sigma_R(\rho)- \frac{1}{2}m^2_\sigma\sigma^2 +
\frac{1}{2}m^2_\omega\omega_0^2 
\nonumber\\
+& \frac{\gamma}{6\pi^2}\int_0^{\infty}\dfrac{dk\,k^4}{(k^2 + 
{M^*}^2)^{1/2}}\left[n(k,T,\mu^*)+\bar{n}(k,T,\mu^*)\right],
\nonumber\\
\label{pressuredd}
\end{align}
with the rearrangement term defined as 
\begin{eqnarray}
\Sigma_R(\rho)=\frac{\partial\Gamma_\omega}{\partial\rho}\omega_0\rho
-\frac{\partial\Gamma_\sigma}{\partial\rho}\sigma\rho_s.
\end{eqnarray}
The mean-fields $\sigma$ and $\omega_0$ are given by
\begin{eqnarray}
\sigma = \frac{\Gamma_\sigma(\rho)}{m_\sigma^2}\rho_s,\quad\mbox{and}\quad
\omega_0 = \frac{\Gamma_\omega(\rho)}{m_\omega^2}\rho,
\label{mfdd}
\end{eqnarray}
with the functional forms of $\rho$ and $\rho_s$ given as in the nonlinear model, 
Eqs.~(\ref{rho})-(\ref{rhos}), with the same distributions functions of 
Eq.~(\ref{fermi-dirac}), and the same form for the effective energy $E^*$. The effective 
nucleon mass and chemical potential are now given, respectively, by 
$M^*=M-\Gamma_\sigma(\rho)\sigma$, and $\mu^*=\mu-\Gamma_\omega(\rho)\omega_0 - 
\Sigma_R(\rho)$.

\section{Results}
The necessary conditions used in the calculation of the critical point
are given by the following expressions:
\begin{eqnarray}
P_c=P(\rho_c , T_c),\quad 
\frac{\partial P}{\partial\rho}\bigg|_{\rho_c , T_c}=0,\quad
\frac{\partial^2 P}{\partial\rho^2}\bigg|_{\rho_c , T_c}=0,\quad
\label{conditions}
\end{eqnarray}
where $P_c$, $\rho_c$ and $T_c$ are, respectively, the critical pressure, density and 
temperature.

The critical parameters $P_c$, $\rho_c$, and $T_c$ are then obtained for each of the 
CRMF parametrizations. The results can be seen in Table~\ref{tabcritical}.
\begin{table}[!htb]
\scriptsize
\caption{Critical values for Consistent RMF models}
\centering
\begin{tabular}{l|c|c|c|c}
\hline\hline
Model & Ref.  & $T_{\rm c}$ (MeV)  & $\rho_c$ (fm$^{-3}$)  & $P_c$~(MeV/fm$^3$)\\
\hline
BKA20  & \cite{PRC81-034323}  & $14.92$ & $0.0458$ & $0.209$  \\ 
BKA22  & \cite{PRC81-034323}  & $13.91$ & $0.0442$ & $0.178$ \\
BKA24  & \cite{PRC81-034323}  & $13.83$ & $0.0450$ & $0.177$ \\ 
BSR8  & \cite{PRC76-045801}  & $14.17$ & $0.0440$ & $0.185$ \\ 
BSR9    & \cite{PRC76-045801}  & $14.11$ & $0.0450$ & $0.185$  \\
BSR10  & \cite{PRC76-045801}  & $13.90$ & $0.0439$ & $0.176$  \\ 
BSR11  & \cite{PRC76-045801}  & $14.00$ & $0.0442$ & $0.179$  \\ 
BSR12  & \cite{PRC76-045801}  & $14.15$ & $0.0448$ & $0.185$  \\ 
BSR15  & \cite{PRC76-045801}  & $14.53$ & $0.0456$ & $0.199$ \\ 
BSR16  & \cite{PRC76-045801}  & $14.44$ & $0.0454$ & $0.196$  \\ 
BSR17  & \cite{PRC76-045801}  & $14.32$ & $0.0451$ & $0.191$  \\ 
BSR18  & \cite{PRC76-045801}  & $14.25$ & $0.0451$ & $0.189$  \\ 
BSR19  & \cite{PRC76-045801}  & $14.28$ & $0.0451$ & $0.190$ \\ 
BSR20  & \cite{PRC76-045801}  & $14.41$ & $0.0464$ & $0.197$  \\ 
FSU-III  & \cite{PRC85-024302}  & $14.75$ & $0.0461$ & $0.205$  \\ 
FSU-IV  & \cite{PRC85-024302}  & $14.75$ & $0.0461$ & $0.205$  \\ 
FSUGold  & \cite{PRL95-122501}  & $14.75$ & $0.0461$ & $0.205$ \\ 
FSUGold4  & \cite{NPA778-10}  & $14.80$ & $0.0456$ & $0.204$ \\ 
FSUGZ03  & \cite{PRC74-034323}  & $14.11$ & $0.0450$ & $0.185$  \\ 
FSUGZ06  & \cite{PRC74-034323}  & $14.44$ & $0.0454$ & $0.196$ \\ 
IU-FSU  & \cite{PRC82-055803}  & $14.49$ & $0.0457$ & $0.196$  \\ 
G2*  & \cite{PRC74-045806}  & $14.38$ & $0.0468$ & $0.192$  \\ 
Z271s2  & \cite{PRC82-055803}  & $17.97$ & $0.0509$ & $0.303$  \\ 
Z271s3  & \cite{PRC82-055803}  & $17.97$ & $0.0509$ & $0.303$ \\ 
Z271s4  & \cite{PRC82-055803}  & $17.97$ & $0.0509$ & $0.303$  \\ 
Z271s5  & \cite{PRC82-055803}  & $17.97$ & $0.0509$ & $0.303$  \\ 
Z271s6  & \cite{PRC82-055803}  & $17.97$ & $0.0509$ & $0.303$  \\ 
Z271v4  & \cite{PRC82-055803}  & $17.97$ & $0.0509$ & $0.303$  \\ 
Z271v5  & \cite{PRC82-055803}  & $17.97$ & $0.0509$ & $0.303$ \\ 
Z271v6  & \cite{PRC82-055803}  & $17.97$ & $0.0509$ & $0.303$  \\ 
DD-F  & \cite{PRC74-035802}  & $15.24$ & $0.0505$ & $0.245$  \\ 
TW99  & \cite{NPA656-331}  & $15.17$ & $0.0509$ & $0.241$  \\ 
DDH$\delta$  & \cite{NPA732-24}  & $15.17$ & $0.0509$ & $0.241$  \\ 
DD-ME$\delta$  & \cite{PRC84-054309} & $15.32$ & $0.0491$ & $0.235$ \\ 
\hline \hline
\end{tabular}
\label{tabcritical}
\end{table}

The results we have computed can be compared with the ones obtained from eight experimental data 
Refs.~\cite{karn1,natowitz,karn2,karn3,karn4,karn5,elliott}. In Table~\ref{tabexp} we show a brief 
compilation of these results. In \cite{elliott}, the authors estimate 
not only the value for $T_c=17.9\pm0.4$~MeV, but also for 
$P_c=0.31\pm0.07$~MeV/fm$^3$, and $\rho_c=0.06\pm0.01$~fm$^{-3}$, all of them related to symmetric 
nuclear matter.

By first analyzing the critical temperature, we can see that only the family Z271 (that 
encompasses all 8 related parametrizations), presents $T_c$ compatible with five 
of the eight experimental points, including the more recent
one~\cite{elliott}. The 
density-dependent models present the critical temperature inside the range of $15 
\leq T_c \leq 19$~MeV proposed by \cite{karn3}. The other critical parameters of 
Ref.~\cite{elliott}, namely, pressure and density, are also compatible
with the ones computed for the Z271 family. The density  dependent family also agrees with this experiment. 

\begin{table}[!htb]
\scriptsize
\caption{Summary of experimental data of
Refs.~\cite{karn1,natowitz,karn2,karn3,karn4,karn5,elliott}}
\centering
\begin{tabular}{c|c|c|c}
\hline\hline
Reference & $T_{\rm c}$ (MeV)  & $\rho_c$ (fm$^{-3}$)  & $P_c$~(MeV/fm$^3$)\\
\hline
\cite{karn1}    & $19\pm 3$   & -  & -   \\
\cite{natowitz} & $16.60\pm 0.89$  &  -  & - \\
\cite{karn2}    & $20\pm 3$    & -  & -    \\
\cite{karn3}    & $17\pm 2$    & -  & -      \\
\cite{karn4}    & $\geq 18$         & -  & -  \\
\cite{karn5}    & $19.5\pm 1.2 / 16.5 \pm 1.0$  & -  & -   \\
\cite{elliott} &  $17.9\pm 0.4$  &  $0.06\pm 0.01$   &  $0.31\pm 0.07$   \\
\hline \hline
\end{tabular}
\label{tabexp}
\end{table}

If we look at the structure of Eq.~(\ref{pnl}), we can understand this agreement with the 
experimental data based on the only term that distinguish such model from the 
$\sigma^3 - \sigma^4$ one, which is those containing the $C$ constant. In this case $C\neq0$.

In the case of the density-dependent model, we can think of a similar structure, since the nonlinear 
behavior of the $\sigma$ field can be represented somehow in the thermodynamical 
quantities, by the density-dependent constant $\Gamma_\sigma(\rho)$. The same occurs with 
the $\omega_0$ field, i.e., the strength of the repulsive interaction is also a 
density-dependent quantity, $\Gamma_\omega(\rho)$. 

\begin{figure}[!htb]
\centering
\includegraphics[scale=0.35]{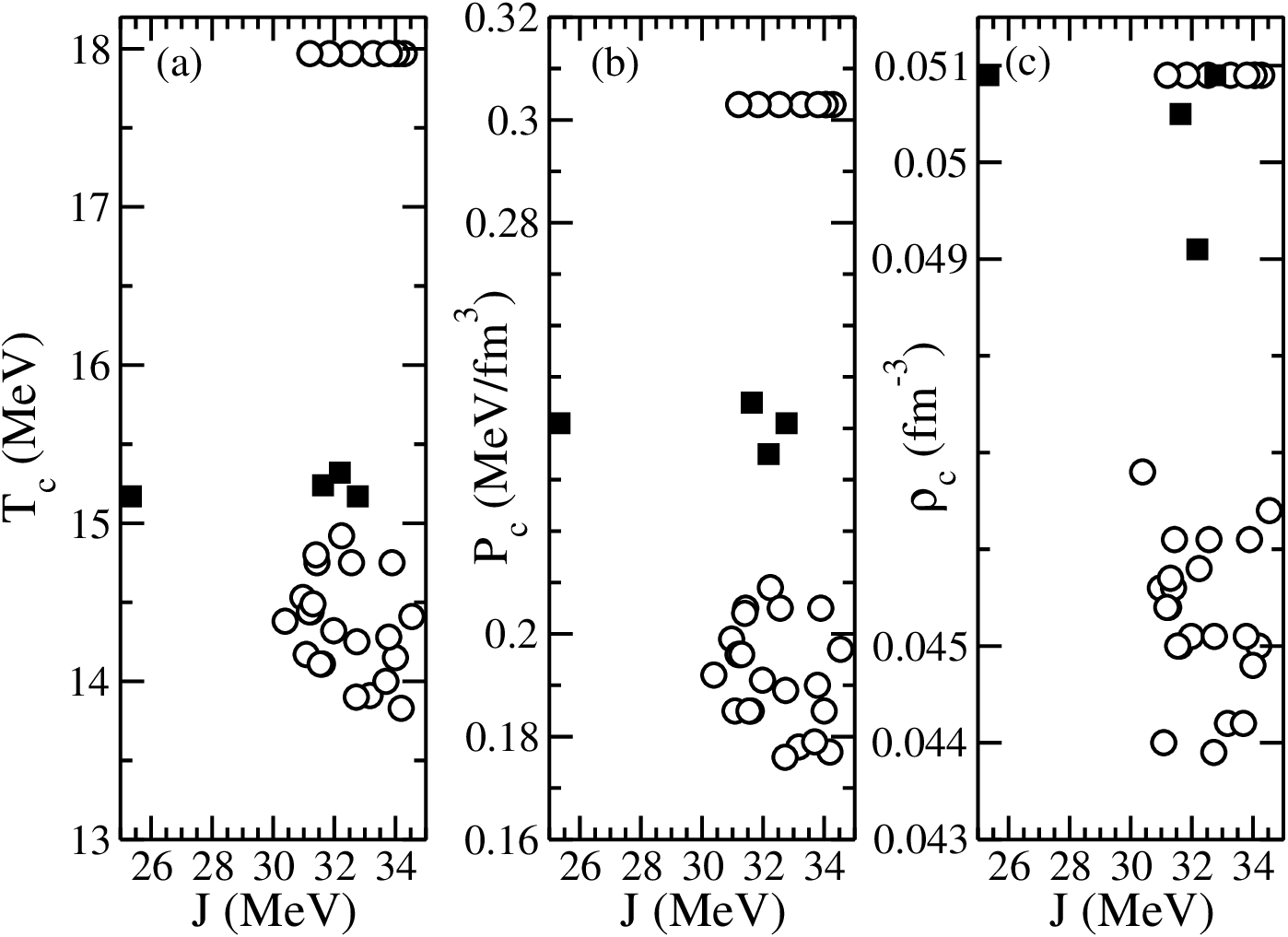}
\vspace{-0.2cm}
\caption{Critical (a) temperature, (b) pressure, and (c) density of CRMF parametrizations versus 
symmetry energy at saturation density. Circles: nonlinear model. Squares: density dependent model.} 
\label{fig1}
\end{figure}

\begin{figure}[!htb]
\centering
\includegraphics[scale=0.35]{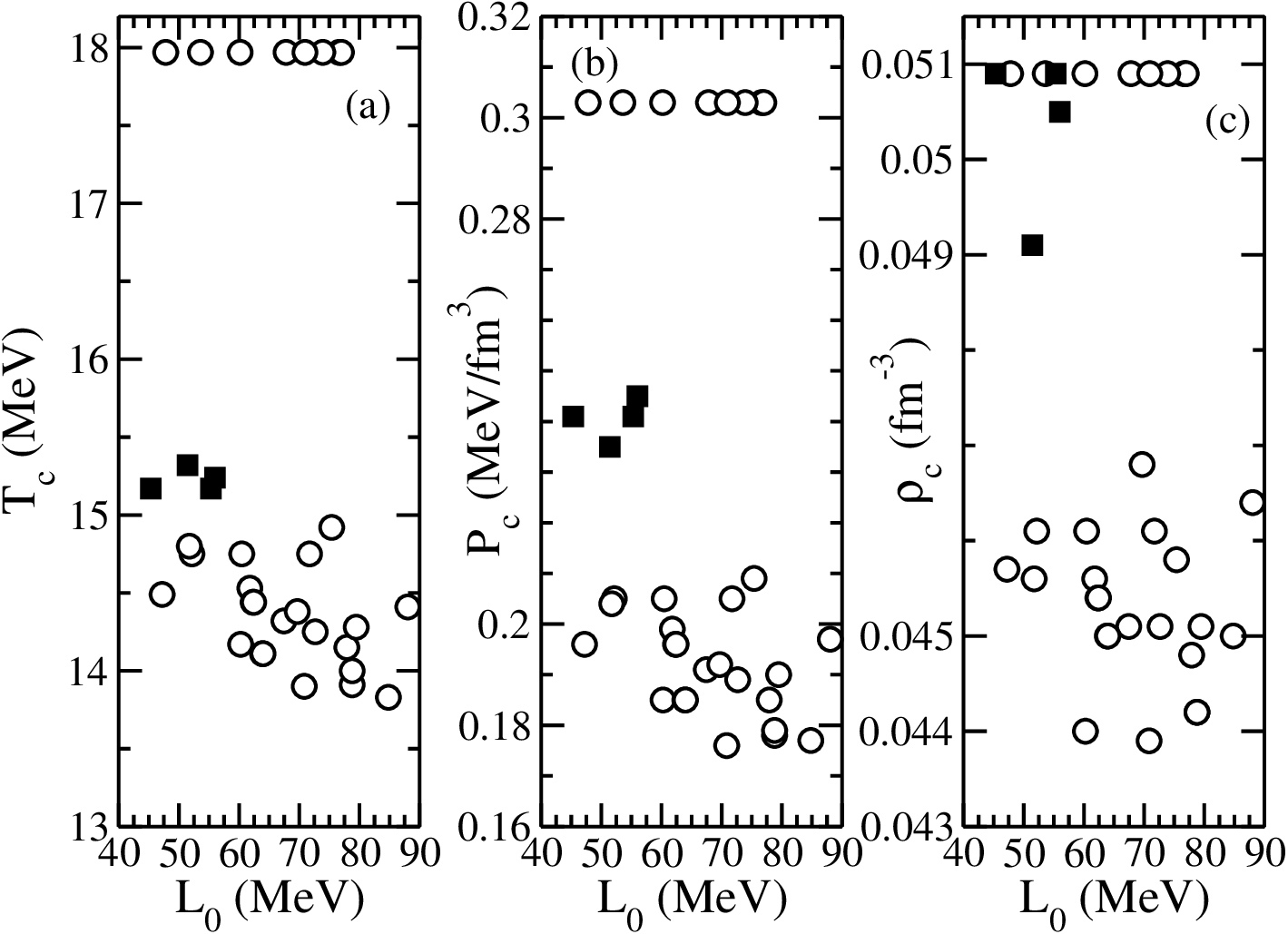}
\vspace{-0.2cm}
\caption{Critical (a) temperature, (b) pressure, and (c) density of CRMF parametrizations 
versus the slope of symmetry energy at saturation density. Circles: nonlinear model. Squares: 
density dependent model.} 
\label{fig2}
\end{figure}

\begin{figure}[!htb]
\centering
\includegraphics[scale=0.35]{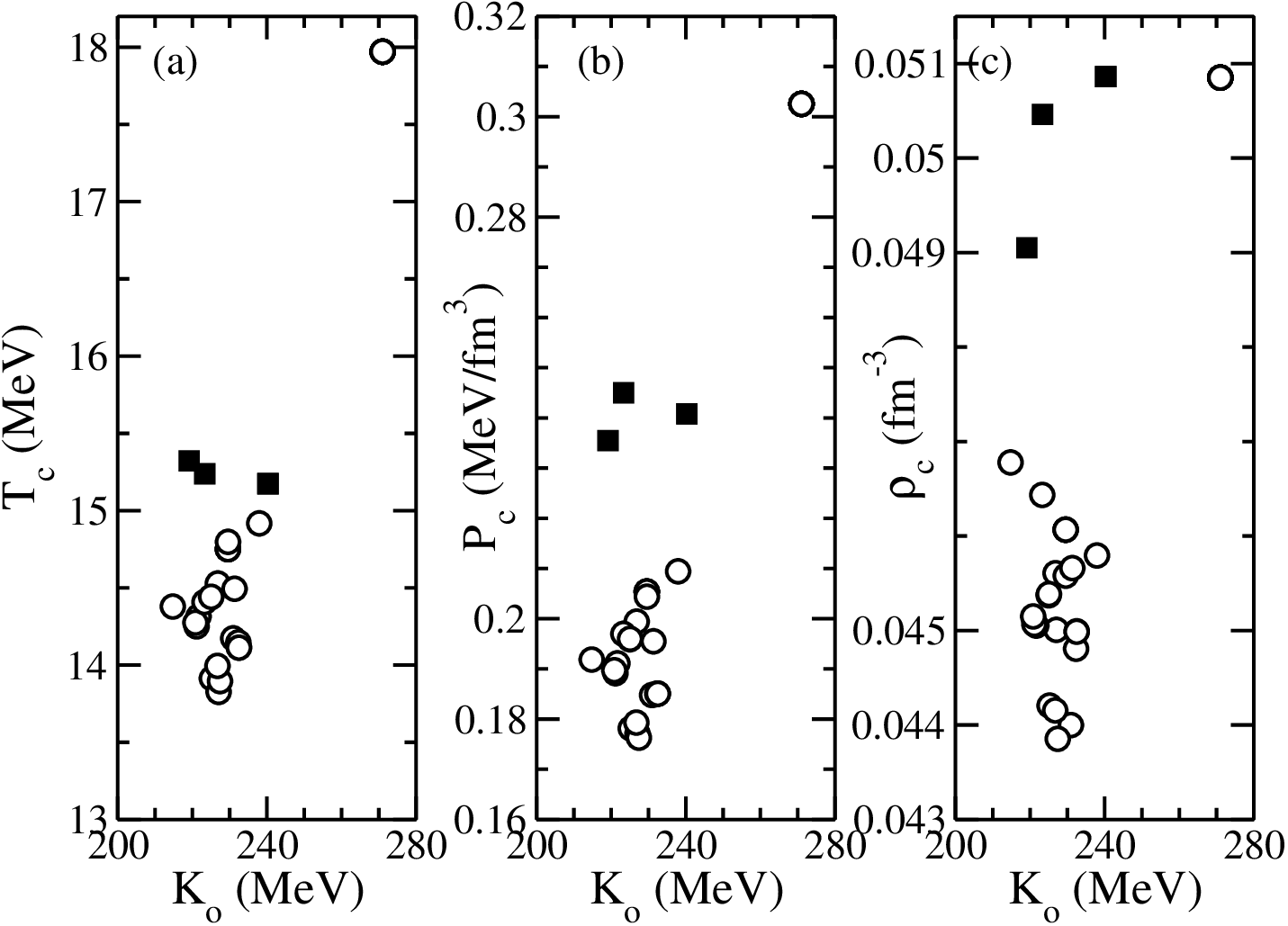}
\vspace{-0.2cm}
\caption{Critical (a) temperature, (b) pressure, and (c) density of CRMF parametrizations 
versus incompressibility at saturation density. Circles: nonlinear model. Squares: density 
dependent 
model.} 
\label{fig3}
\end{figure}

We have also tried to verify if there are correlations between the critical parameters and the observables of 
nuclear matter at zero temperature and at the saturation density. We investigate possible 
correlations between $T_c$, $P_c$ and $\rho_c$ with the symmetry energy, its slope and 
incompressibility. The results are shown in Figs.~\ref{fig1}, \ref{fig2}, and~\ref{fig3}, 
respectively.

Note that for the symmetry energy and its slope, Figs.~\ref{fig1} and
\ref{fig2},  there are no indications of possible correlations. A
unique pattern for the nonlinear and density-dependent models are not seen.
However, the picture changes when we look at the incompressibility ($K_0)$. From Fig.~\ref{fig3},
one can observe an increasing behavior of $T_c$, $P_c$ and $\rho_c$ as $K_0$ increases. 

\section{Summary}

In this work, we present the results obtained in the calculation of the critical parameters: 
temperature, pressure, and density in symmetric nuclear matter. In our analysis, we 
verified that the nonlinear models, whose parameterizations were grouped in the family 
Z271, show a good agreement with the experimental data~\cite{elliott} for all critical parameters 
analyzed. The density-dependent family also shows an agreement with
the data given in ~\cite{elliott}  for the  pressure and
density. Concerning $T_c$, the agreement 
is only found with data presented in  Ref.~\cite{karn3}.

In the search for possible correlations, we can see that the 
incompressibility at zero temperature and at saturation density show a
clear increasing behavior with the critical parameters analyzed. 
The same does not occur with the symmetry energy and its slope.

\section*{Acknowledgments} 

This work was partially supported by Conselho Nacional de Desenvolvimento Cient\'ifico e 
Tecnol\'ogico (CNPq), Brazil under grants 301155/2017-8 and
310242/2017-7. This work is also a part of the project CNPq-INCT-FNA Proc. No. 464898/2014-5.


\begin{thebibliography}{99}

\bibitem{prc-critical} O. Louren\c co, M. Dutra, and D. P. Menezes, Phy. Rev. C {\bf 95}, 065212 
(2017)

\bibitem{vdw5} J. B. Silva, O. Louren\c{c}o, A. Delfino, J. S. S\'a Martins, M. Dutra, 
Phys. Lett. B {\bf 664} 246, (2008).

\bibitem{avancini2006} S.S. Avancini, L. Brito, Ph. Chomaz, D.P. Menezes and C. Provid\^encia, 
Phys. Rev. C {\bf 74}, 024317 (2006).

\bibitem{rmf} M. Dutra, O. Louren\c{c}o, S. S. Avancini, B. V. Carlson, A. Delfino, D. P. 
Menezes, C. Provid\^encia, S. Typel, and J. R. Stone,  Phys. Rev. C {\bf 90}, 055203 (2014).

\bibitem{stars} M. Dutra, O. Louren\c{c}o, and D. P. Menezes, Phys. Rev. C {\bf 93}, 025806 (2016); 
Erratum:  Phys. Rev. C {\bf 94}, 049901(E) (2016).

\bibitem{NPA656-331} S. Typel and H. H. Wolter. Nucl. Phys. A {\bf 656}, 331 (1999).

\bibitem{PRC81-034323} B. K. Agrawal. Phys. Rev. C {\bf 81}, 034323 (2010).

\bibitem{PRC76-045801} S. K. Dhiman. R. Kumar. and B. K. Agrawal. Phys. Rev. C {\bf 76}, 045801 
(2007).

\bibitem{PRC85-024302} B.-J. Cai. L.-W. Chen. Phys. Rev. C {\bf 85}, 024302 (2012).

\bibitem{PRL95-122501} B. G. Todd-Rutel and J. Piekarewicz. Phys. Rev. Lett. {\bf 95}, 122501 
(2005).

\bibitem{NPA778-10} J. Piekarewicz and S. P. Weppner. Nucl. Phys. A {\bf 778}, 10 (2006).

\bibitem{PRC74-034323} R. Kumar. B. K. Agrawal. and S. K. Dhiman. Phys. Rev. C {\bf 74}, 034323 
(2006).

\bibitem{PRC74-045806} A. Sulaksono and T. Mart. Phys. Rev. C {\bf 74}, 045806 (2006).

\bibitem{PRC82-055803} F. J. Fattoyev. C. J. Horowitz. J. Piekarewicz. and G. Shen. Phys. Rev. C 
{\bf 82}, 055803 (2010).

\bibitem{PRC74-035802} T. Kl\"ahn. {\it et al.}. Phys. Rev. C {\bf 74}, 035802 (2006).

\bibitem{NPA732-24} T. Gaitanos. M. Di Toro. S. Typel. V. Baran. C. Fuchs. V. Greco. and H. H. 
Wolter. Nucl. Phys. A {\bf 732}, 24 (2004).

\bibitem{PRC84-054309} X. Roca-Maza. X. Vi\~nas. M. Centelles. P. Ring. and P. Schuck Phys. Rev. C 
{\bf 84}, 054309 (2011).

\bibitem{karn1} V. A. Karnaukhov, Phys. At. Nucl. {\bf 60}, 1625 (1997).

\bibitem{natowitz}  J. B. Natowitz, K. Hagel, Y. Ma, M. Murray, L. Qin, R. Wada, and J. Wang, Phys. 
Rev. Lett. {\bf 89}, 212701 (2002).

\bibitem{karn2} V. A. Karnaukhov, et al., Phys. Rev. C {\bf 67}, 011601(R) (2003).

\bibitem{karn3} V. A. Karnaukhov et al., Nucl. Phys. A {\bf 734}, 520 (2004).

\bibitem{karn4} V. A. Karnaukhov et al., Nucl. Phys. A {\bf 780}, 91 (2006).

\bibitem{karn5} V. A. Karnaukhov, Phys. At. Nucl. {\bf 71}, 2067 (2008).

\bibitem{elliott} J. B. Elliott, P. T. Lake, L. G. Moretto, and L. Phair, Phys. Rev. C {\bf 87}, 
054622 (2013).

\end{thebibliography}
\end{document}